\documentclass[twocolumn]{aastex61}

\received{June 1, 2016}
\revised{Mar 6, 2017}
\accepted{Mar 8, 2017}
\submitjournal{ApJ}

\begin{document}

\title{A Curved Magnetic field in the ring-like shell of bubble N4}

\correspondingauthor{Zhiwei Chen}
\email{zwchen@pmo.ac.cn}

\author[0000-0003-0849-0692]{Zhiwei Chen}
\affil{Purple Mountain Observatory $\&$ Key Laboratory for Radio 
Astronomy, Chinese Academy of Sciences, 2 West Beijing Road, 210008 Nanjing, China} 

\author{Zhibo Jiang}
\affil{Purple Mountain Observatory $\&$ Key Laboratory for Radio 
Astronomy, Chinese Academy of Sciences, 2 West Beijing Road, 210008 Nanjing, China}

\author{Motohide Tamura}
\affil{Department of Astronomy, Graduate School of Science, The University of Tokyo, 7-3-1 Hongo, Bunkyo-ku, Tokyo 113-0033, Japan}
\affil{National Astronomical Observatory of Japan, 2-21-1 Osawa, Mitaka, Tokyo 181-8588, Japan}

\author{Jungmi Kwon}
\affil{Institute of Space and Astronautical Science, Japan Aerospace Exploration Agency, 3-1-1 Yoshinodai, Chuo-ku, Sagamihara, Kanagawa 252-5210, Japan}

\author{A. Roman-Lopes}
\affil{Department of Physics and Astronomy, Universidad de La Serena, Av. Juan Cisternas, 1200 La Serena, Chile}

\begin{abstract}
We report the detection of a curved magnetic field in the ring-like shell of the bubble N4, derived from near-infrared polarization of reddened diskless stars located behind this bubble. The magnetic field in the shell is curved and parallel to the ring-like shell, and its strength is estimated to be $\sim120\,\mu$G in the plane of the sky. The magnetic field strength in the shell is significantly enhanced compared to the local field strength. We calculate the mass-to-flux ratio for the submillimeter clumps in the shell and find that they are all magnetically subcritical. Our results demonstrate that the magnetic field strengthens as the interstellar medium is compressed into a shell, and suggest that the magnetic field has the potential to hinder star formation triggered by \ion{H}{2} region expansion.
\end{abstract}


\section{Introduction} \label{sec:intro}

Parsec-scale bubbles are created as stellar winds and/or \ion{H}{2} region ionization fronts from massive stars expand into the interstellar 
medium \citep[ISM;][]{2006ApJ...649..759C}. The expansion of a bubble sweeps up ISM into a shell of enhanced density, 
which could result in fragmentation into compact star-forming cores, the so-called ``collect and collapse'' model of triggered star 
formation \citep{1998ASPC..148..150E}. \citet{2010A&A...523A...6D} used multi-wavelength public data to resolve the spatial distributions 
of cold dust condensations in the borders of 65 bubbles and found that 40\% of them are good candidates for the ``collect and collapse'' process. 
Nevertheless, these good candidates need follow-up studies to determine their physical conditions and to verify if star formation is triggered by 
the ``collect and collapse'' process. \citet{2010ApJ...709..791B} mapped seven bubbles with dense gas tracers (CO\,$J=3-2$ and HCO$^+$\,$J=4-3$ lines), 
and derived column densities which are not high enough for initiating the ``collect and collapse'' process of induced star formation. They proposed 
that bubbles expanding into flattened molecular clouds might not be able to collect sufficient gas into shells and would be unable to trigger the formation of 
new stars. However, the role of the magnetic field is not considered in any of the above studies.

The effects of magnetic fields in the expansion of bubbles have been studied in magnetohydrodynamic simulations. These show that the magnetic field gets stronger in the 
shell as the ISM is compressed and the field greatly reduces the strength of shocks and the density contrast in directions perpendicular to 
the magnetic field. The strong, ordered magnetic field in the shell hinders the formation of thin-shell instabilities and reduces the efficiency of star formation triggered by massive stars \citep{2007ApJ...671..518K,2015A&A...584A..49V}. Another result of magnetic fields influencing the expansion of bubbles is that the elongation direction of an elliptical bubble is parallel to the magnetic field direction. Although several aspects, such as a non-uniform ISM and the inclination angles can affect the projected elongation of a bubble in a realistic ISM, this phenomenon can be investigated observationally. \citet{2012ApJ...760..150P} found that the superthermal \ion{H}{2} region--driven bubbles are preferentially aligned with the average orientation of the Galactic magnetic field in the disk, and subthermal \ion{H}{2} region--driven bubbles are consistent with random alignments. This observational result implies that magnetic fields play important roles in the early evolution of \ion{H}{2} region--driven bubbles. However, a detailed observational study of the influence of magnetic fields on bubbles has not been performed to date. 

In this paper we report measurements of the magnetic field for the bubble N4 and the associated star-forming clump N4W. The bubble N4 is a ring-like shell structure enclosing the \ion{H}{2} region G11.898+0.747 identified by \citet{1989ApJS...71..469L}. The molecular gas of N4
shows a systematic velocity $V_\mathrm{LSR}\sim25$\,km/s \citep{2013RAA....13..921L}, which suggests a most probable distance of $2.80\pm0.30$\,kpc, according to the parallax-based distance estimator \citep{2016ApJ...823...77R}. Several papers have sought evidence for triggered star formation in the bubble N4 \citep[e.g.][]{2010A&A...523A...6D,2010ApJ...716.1478W,2016ApJ...818...95L,2016AJ....152..117Y}, and all concluded there is no such formation in N4. At the same $V_\mathrm{LSR}$ of the bubble N4, N4W is found to harbor several intermediate-mass Class I/II objects and one cold dense molecular core \citep[][hereinafter, Paper I]{2016ApJ...822..114C}.
The current paper makes use of data taken using
SIRPOL, mounted on the Infrared Survey Facility 1.4\,m telescope at the South African
Astronomical Observatory, during the nights of 2013 July 7--9. The instrument performance, observational information, weather conditions, and data 
reduction process were previously described in Paper I.

\begin{figure}[ht!]
\figurenum{1}
\centering
  \plotone{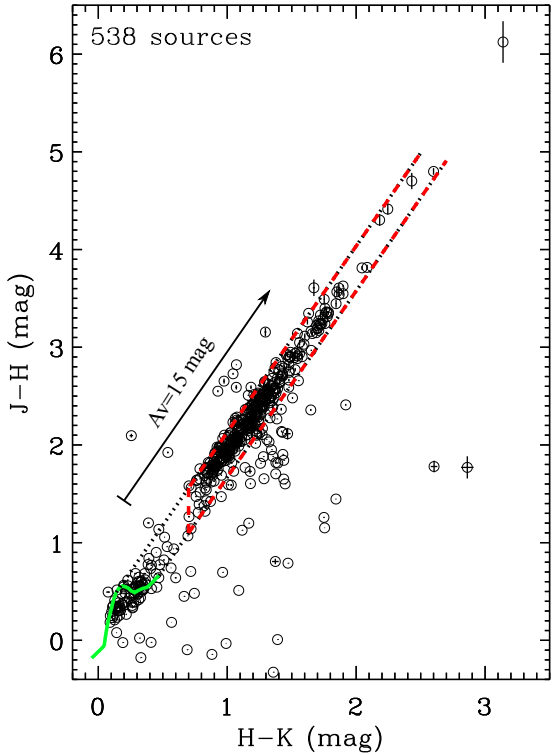}
 \caption{ $J-H$ versus $H-K$ color-color diagram of the 538 sources with both UKIDSS/GPS DR6 photometry and SIRPOL $H$-band polarization. 
The red dashed lines are reddening vectors. The main-sequence intrinsic colors (highlighted in green) are adopted from \citet{2013ApJS..208....9P}.}
 \label{Fig:ccd}
\end{figure}

% \begin{figure*}[ht!]
% \figurenum{2}
%   \plotone{gas+components}
%  \caption{Extinction map derived from the $^{12}$CO\,$J=1-0$ line emission in bubble N4. Blue contours correspond to extinction values $A_V=2,3,4$\,magnitudes derived from $^{12}$CO\,$J=1-0$ line emission of the closer $V_\mathrm{LSR}\sim15\,\mathrm{km/s}$ molecular cloud. The $H$-band polarization values of the foreground candidates are shown by the white lines, with the lengths being proportional to the polarization degree values. Also a scale length of 10\% is denoted by the red line. The red dashed line in the upper right corner indicates the orientation of the Galactic plane. }
%  \label{Fig:gas}
% \end{figure*}

\begin{figure}[ht!]
\figurenum{2}
\centering
%\epsscale{1}
\plotone{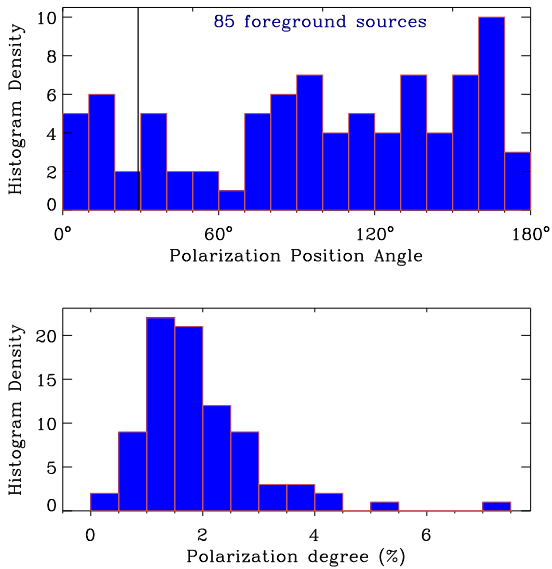}
\caption{Top panel: Distribution of $\theta_H$ of the foreground stars. The vertical line represents the Galactic plane orientation of $29\degr$. Bottom panel: Distribution of degree of polarization $P_H$ of the foreground stars. }
\label{Fig:fore}

\end{figure}

\begin{figure*}[ht]
\figurenum{3}
  \plotone{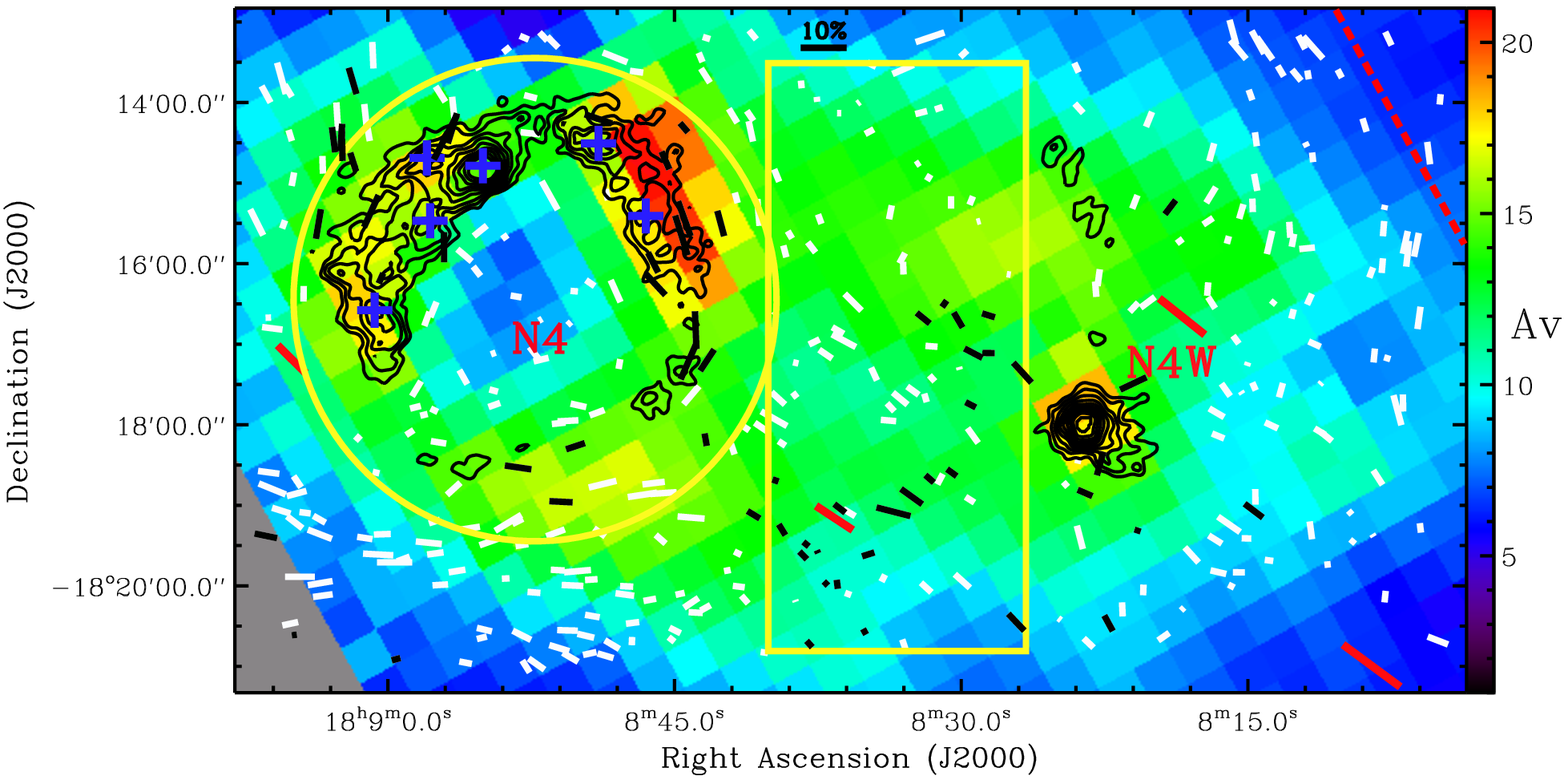}
 \caption{$H$-band polarization of background stars (white lines) overlaid on the extinction map derived from the $^{12}$CO\,$J=1-0$ data from \citet{2013RAA....13..921L}. Very red background stars ($A_V>20$\,mag) are highlighted as black lines. The red lines are 353\,GHz dust polarization calculated by the $5\times5$ rebinning of the $Planck$ data with an original pixel 
 scale of $1\arcmin$. The black contours, of 0.26 to 1.16 mJy/beam with an interval 
 of 0.1 mJy/beam, are $870\,\mu$m dust continuum emission from ATLASGAL \citep{2009A&A...504..415S}. Blue pluses denote the six submillimeter clumps identified by \citet{2014A&A...565A..75C}. 
 The yellow circle roughly outlines the size of the bubble N4, and the yellow box covering N4W is defined as the control region for measuring the mean local magnetic field. The red dashed line denotes the Galactic plane orientation.}
\label{Fig:allp}
\end{figure*}

\section{Polarimetry of point sources} \label{sec:obs}
The average full width at half maximum (FWHM) of point sources in the $J$-, $H$-, and $K_s$-bands is $2\farcs0$, $1\farcs9$, and $1\farcs8$, respectively.  We apply a fixed aperture radius of $1\farcs35$ (3 pixels) to conduct aperture photometry on the reduced $JHK_s$ images at 
each wave-plate angle. To secure precise photometry, we only consider data from objects fulfilling two conditions: 1) the object's FWHM is between $1\farcs4$ and $2\farcs7$, not largely different from the mean FWHM, and 2) the object is not blended with another at the spatial resolution of the SIRPOL data. The integrated detector counts of these point sources are used to calculate the Stokes parameters $I$, $U$, $Q$ (Equation (1) of Paper I), which 
together provide the degree of polarization $P$ and the polarization position angle $\theta$ according to Equation (2) of Paper I. The position angle $\theta$ 
is measured anticlockwise from north to east. The error of $P$ can be obtained by $\delta P=\frac{\sqrt{2}}{2}~\delta I/I$, where $\delta I$ is 
the photometric error. The error of $\theta$ is $\delta \theta=28\fdg6~\delta P/P$; thus larger $P/\delta P$ corresponds to smaller $\delta \theta$. In the following analysis, we only consider point sources satisfying $P\geqslant 3\,\delta P$.

\section{Classifying reddened background stars}\label{sec:ccd}

The SIRPOL $JHK_s$ images are too shallow to penetrate through the dense regions in in N4 and N4W. In order to circumvent this problem, we retrieve much deeper $JHK$ photometric data from the DR6 release of the UKIDSS/GPS survey\footnote{http://www.ukidss.org/surveys/gps/gps.html}. 
We cross-match the DR6 release of UKIDSS/GPS and the $H$-band polarization data. Applying a maximum separation of $0\farcs5$ in cross-match, we derive a sample of 538 point sources with both UKIDSS/GPS $JHK_s$ photometry and $H$-band polarization. The distribution of these 538 point sources in the $H-K$ versus $J-H$ diagram is shown in Fig.~\ref{Fig:ccd}. There are 441 point sources located between the two parallel reddening vectors (dotted lines in Fig.~\ref{Fig:ccd}). These are diskless stars whose extinctions are due to interstellar dust along the line of sight (LOS). Therefore, the polarization of these diskless stars is mostly dominated by interstellar polarization originating from the dichroic extinction of nonspherical dust grains with short axes aligned along the interstellar magnetic field \citep[see the review by][]{2015ARA&A..53..501A}. Thus, the polarization of these diskless stars is expected to be parallel to the interstellar magnetic field. Indeed, the observed polarizations of these stars exhibit the mean direction of the interstellar magnetic field averaged along the LOS. In order to trace the magnetic field in the bubble N4, it is necessary to identify stars whose polarization is caused by dichroic extinction by dust grains in the the bubble.

The obvious gap at around $H-K=0.6$\,mag in Fig.~\ref{Fig:ccd} naturally splits the 441 diskless stars into low-extinction and high-extinction groups. The extinction of diskless stars is estimated by applying the conversion \footnote{In the conversion from the $H-K$ color to $A_V$, a transformation from the UKIRT photometric system to the 2MASS photometric system is applied \citep{2001AJ....121.2851C}. The conversion factor of 14.7 is derived from the recent near-infrared extinction law $A_\lambda\propto\lambda^{-1.95}$ \citep{2014ApJ...788L..12W} and a reasonable assumption $A_J=0.29$ when $R_V=3.1$. } equation $A_V=14.7~([H-K]_{obs}-0.2)$ to their $H-K$ colors. The low-extinction group have less dust extinction and show a smaller extinction range ($A_V<6$\,mag), indicating they are foreground objects. The high-extinction group show a much larger extinction range ($7<A_V<50$\,mag), most likely contributed by the ISM in the bubble N4. We regard the reddened diskless stars ($H-K>0.7$\,mag) as the background stars. The $H$-band polarization data of these 356 background stars
are listed in Table 1. 

Although the foreground objects have lower extinctions, a number of them exhibit $H$-band polarization at high significance levels ($S/N>3$). Fig.~\ref{Fig:fore} presents the distributions of the polarization position angles and degrees of polarization for these foreground stars. Most of these foreground stars show a degree of polarization $P_H<2\%$, with two exceptions that show extremely large values ($P_H>5\%$). On the other hand, the position angle distribution of these foreground stars is roughly flat. A one-sided Kolmogorov--Smirnov test shows that the probability that the foreground $\theta_H$ distribution was drawn from a random distribution is 0.14, much higher than the typical significance level of 5\%. The result of the Kolmogorov--Smirnov test indicates that the foreground $\theta_H$ distribution is very close to a random distribution. 

In Fig.~\ref{Fig:allp} we overlaid the $H$-band polarization of the 356 background stars on the extinction map of the bubble N4 \footnote{In the transformation from $N(\mathrm{H}_2)$ to $A_V$, the equation $A_V=9.0\times10^{-22}~\mathrm{cm}^{2}~N(\mathrm{H}_2)~\mathrm{mag}$ from \citet{2009MNRAS.400.2050G} is used.}. The polarization vectors of the background stars show a certain degree of alignment in smaller areas. For instance, the polarization vectors are oriented at about $90\degr$ in the area just below the bubble N4; in the area between the bubble N4 and the star-forming clump N4W, we find the polarization vectors are roughly parallel to the Galactic plane orientation of $29\degr$. The area below the bubble N4 is a region of low-density molecular gas with $A_V\sim7\,$mag, as inferred from the extinction map shown in Fig.~\ref{Fig:allp}. The contribution of the unrelated foreground ISM along the same LOS may be more or less comparable to the local ISM in the bubble N4. Therefore, the polarization direction of background stars in this area is affected by both components. In contrast, the area between N4 and N4W is high-density, so that the reddening of background stars in this area is mostly contributed by the ISM in the bubble N4.

%{The photometries in the $JHK_s$ bands are necessary for classifying highly reddened background stars on the basis of $H-K_s$ versus $J-H$ diagram. We classified 120 point sources as reddened background stars with $JHK_s$ polarizations. Overall 62\% of the 120 reddened background stars show $\theta$ value differences less than $30\degr$ in all three bands. Therefore the polarizations of the 120 reddened background stars are majorly due to the dichroic extinction of interstellar dust. Nevertheless, this small number of reddened background stars are not enough in tracing the magnetic field direction towards the bubble N4. } 
\floattable
\begin{table}[h!]
\centering
\tablenum{1}
\caption{$H$-band Polarimetric Data of Background Stars Behind the Bubble N4}\label{tbl:pol}
\begin{tabular}{cccccc}
\tablewidth{0pt}
\hline
\hline
 R.A. & Decl. & $P_H$ & $\delta P_H$ & $\theta_H$ & $\delta \theta_H$ \\
 (J2000) & (J2000) & \% & \% & $\degr$ & $\degr$ \\ 
\hline
272.020601 &	-18.232673 &	2.1 &	0.1 &	174 &	1 \\ 
272.021080 &	-18.344659 &	4.5 &	0.1 &	69 &	1 \\
272.023085 &	-18.243796 &	4.6 &	0.6 &	7 &	4 \\
272.024903 &	-18.321352 &	2.0 &	0.4 &	7 &	6 \\
272.025196 &	-18.277893 &	4.6 &	0.5 &	172 &	3 \\
272.025758 &	-18.262770 &	5.0 &	0.2 &	169 &	1 \\
272.027667 &	-18.264612 &	5.3 &	0.5 &	173 &	3 \\
272.028186 &	-18.262282 &	4.8 &	0.3 &	168 &	2 \\
272.028691 &	-18.295514 &	3.5 &	0.2 &	1 &	1 \\
\hline
\end{tabular}
\tablecomments{This table is available in its entirety in machine-readable form.}
\end{table}

\begin{figure}[ht!]
\figurenum{4}
\centering
 \plotone{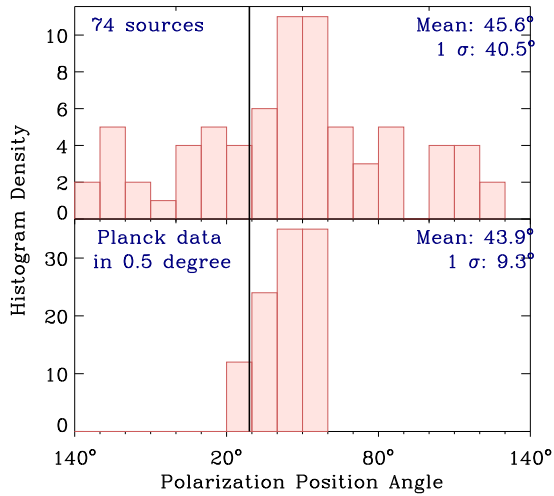}
\caption{Top panel: Distribution of $\theta_H$ for the background stars in the control region.
Bottom panel: Distribution of $\theta+90\degr$ for the Planck 353\,GHz dust polarization within $0.5\degr$ radius. The vertical line represents the Galactic plane orientation at $29\degr$.}
\label{Fig:allPA}
\end{figure}

%Ok, revision/suggestions (from Alex) to this section done!

%\subsection{Cross-match with UKIRT/GPS catalog}

\begin{figure}[ht!]
\figurenum{5}
  \plotone{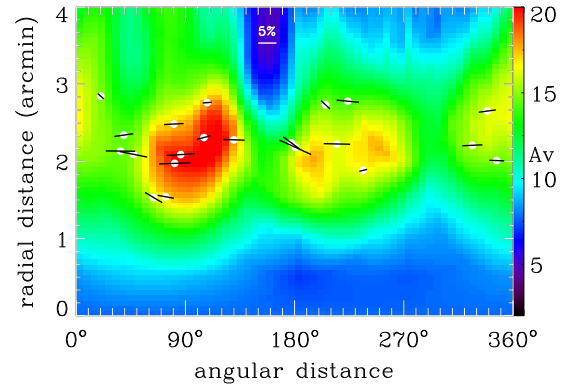}
 \caption{$H$-band polarization vectors of background stars located behind the shell of the bubble N4. The background color image represents the extinction data remapped in polar coordinates. The angular distance is counted anticlockwise from zero (parallel to the Galactic plane) to 360$\degr$. The position angle $\theta$ was corrected to be $\theta=0\degr$ when polarization is locally parallel to the shell. }
 \label{Fig:annulus}

\end{figure}

\begin{figure}
\figurenum{6}
\centering
  \plotone{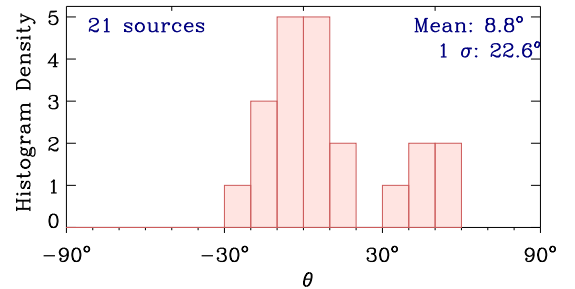}
\caption{Distribution of corrected $\theta$ for the background stars shown in Fig.~\ref{Fig:annulus}}
\label{Fig:annu_PA}
\end{figure}

\floattable
\begin{deluxetable}{ccccccc}
\tablenum{2}
\tablecaption{Physical properties of the dust condensations on the shell of N4}\label{tbl:clump}
\tablewidth{0pt}
\tablehead{
\colhead{ID} & \colhead{R.A.} & \colhead{Decl.} & \colhead{FWHM} & \colhead{Integrated $870\,\mu$m Flux} & \colhead{Mass} & \colhead{Mass-to-flux Ratio} \\
& \colhead{(J2000)} & \colhead{(J2000)} & \colhead{(pc)} & \colhead{(Jy)} & \colhead{($M_\sun$)}} 
\decimalcolnumbers
\startdata
  G011.8937+0.7780 &  18 8 46.5 &  -18 15  24 &   0.45 & 1.23 &  47.6 &  $\lesssim 0.2$ \\
  G011.9116+0.7767 &  18 8 49.0 &  -18 14  30 &   0.41 & 1.18 &  45.6 &  $\lesssim 0.3$ \\
  G011.9191+0.7536 &  18 8 55.0 &  -18 14  47 &   0.37 & 2.42 &  93.6 &  $\lesssim 0.7$ \\
  G011.9145+0.7384 &  18 8 57.8 &  -18 15  28 &   0.37 & 0.91 &  35.2 &  $\lesssim 0.3$ \\
  G011.9261+0.7442 &  18 8 58.0 &  -18 14  41 &   0.37 & 0.89 &  34.4 &  $\lesssim 0.3$ \\
  G011.9039+0.7194 &  18 9  0.7 &  -18 16  35 &   0.53 & 1.98 &  76.6 &  $\lesssim 0.3$ \\
\enddata 
\tablecomments{Assuming the clumps are spherical, with the FWHM as their radius.}
\end{deluxetable}

\section{Magnetic field in the shell of the bubble N4}
The $H$-band polarization map obtained from the stars behind the ring-like shell of the bubble N4 looks curved and follows the arc-like shape of the shell. This is particularly true for the black lines. In contrast, the $H$-band polarizations of background stars surrounding the bubble N4 show the distinct directions mentioned above. To check the orientations against that of the mean local magnetic field, we compare the $\theta_H$ of background stars located between N4 and N4W (defined as the control region, the yellow rectangle in Fig.~\ref{Fig:allp}) and the position angles of the dust polarization at 353\,GHz from the $Planck$ mission\footnote{Based on observations obtained with $Planck$ (http://www.esa.int/Planck), an ESA science mission with instruments and contributions directly funded by ESA member states, NASA, and Canada.}. Fig.~\ref{Fig:allPA} shows the position angle distribution of the control region and the $Planck$ 353\,GHz dust polarization for positions within a $0\fdg5$ radius. The mean local magnetic field direction, quantified by the mean position angle of background stars in the control region, agrees well with that traced by the dust polarization at larger scales. The consistency between them indicates a mean local magnetic field direction $\sim45\degr$, which is nearly parallel to the Galactic plane, oriented at $29\degr$.
   
The orientations of the remaining (non--Galactic plane) polarization vectors closely match the ring-like shell of the bubble N4. Fig.~\ref{Fig:annulus} demonstrates the curvature of the magnetic field along the ring-like shell. The position angles of the background stars, relative to the shell's local orientation, have a mean value of $8\fdg8$ (see Fig.~\ref{Fig:annu_PA}), indicating a very tight correlation between the magnetic field direction and the shell orientation. The curved magnetic field in the shell is the result of magnetically frozen gas being swept up into a shell. This observational phenomenon of a curved magnetic field in a bubble's shell  provides a vital benchmark for studying the effects of magnetic fields when \ion{H}{2} region--driven bubbles expand into the magnetized ISM.

In order to estimate the magnetic field strength in the shell, we use the Chandrsekhar--Fermi method \citep{1953ApJ...118..113C}: $$B_\mathrm{sky}=Q\sqrt{4\pi\rho}\frac{\delta v}{\delta\phi},$$ where $Q$ is the correction factor, $\rho$ is the mass density of the molecular gas, $\delta v$ is the one-dimensional velocity dispersion, and $\delta \phi$ is the $\theta_H$ dispersion of the background stars lying behind the gas shell. The value of $\delta \phi$ is calculated to be $22\fdg8$ after correcting for measurement uncertainties.
When $\delta \phi$ is not large ($<25\degr$), a correction factor $Q\approx 0.5$ yields a reliable field strength estimate \citep{2001ApJ...546..980O}. Then, the above equation can be rewritten into the form
$$B_\mathrm{sky}\approx9.3\sqrt{n_\mathrm{H_2}}\,\frac{\Delta V}{\delta \phi}\, \mu G,$$
 where $\Delta V$ is the line width and $n_\mathrm{H_2}$ is the gas number density.  We calculate $\Delta V=2.0\,\mathrm{km/s}$ and $n_\mathrm{H_2}=2.1\times10^4\,\mathrm{cm}^{-3}$ from the C$^{13}$O\,$J=1-0$ line emission of the bubble N4 \citep{2013RAA....13..921L}. The resulting field strength in the plane of the sky is estimated to be $120\,\mu$G for the gas shell of the bubble N4. 

For the control region, we calculate $\delta \phi=40\fdg5$ from the $\theta_H$ of background stars in the control region. However, the position angle dispersion is quite large, indicating a magnetic field strength lower than that in the bubble's shell.

\section{Discussion and Concluding remarks}
The magnetic field in the shell is stronger than that in the ambient field. In the ``collect and collapse'' scenario, the strength of the magnetic field in the shell is vital to the evolution of the shell. Because the magnetic field becomes strong in the shell, the role of the magnetic field may become more important as well. 
Six dust condensations in the shell of N4 are found in the ATLASGAL dust condensation catalog of the Galactic plane \citep{2014A&A...565A..75C}. Their properties are listed in the Col. (1)-(5) of Table~\ref{tbl:clump}. Given the distance of 2.8\,kpc, the mass of the six clumps is derived by applying the Equation (2) of \citet{2014A&A...565A..75C}\footnote{ We adopt the same assumptions  as \citet{2014A&A...565A..75C}, i.e., a gas-to-dust ratio $R=100$ and dust opacity $\kappa\approx1.85\mathrm{cm}^2\mathrm{g}^{-1}$ at 870\,$\mu$m. The most massive 
clump in the shell, G011.9191+0.7536, presents a rotational temperature of $21.6\pm4.1\,\mathrm{K}$ measured from NH$_3$ lines \citep{2012A&A...544A.146W}.  
We assume this rotational temperature as the dust temperature for all six clumps in the shell, i.e., $T_\mathrm{d}=21.6\,\mathrm{K}$.}. Based on the mean magnetic field strength in the shell calculated above, the critical mass that can be 
maintained by the magnetic field is \citep{1987ARA&A..25...23S}
$$M_{cr}=0.13\frac{\Phi}{G^{1/2}}\approx833 M_\odot\left(\frac{B}{100\,\mu\mathrm{G}}\right)\left(\frac{R}{1\,\mathrm{pc}}\right)^2,$$ where the radius $R$ 
is assumed to be equal to the clump's FWHM. Note that the measured field strength only corresponds to the field component in the plane of the sky; thus it is  a lower limit to the total field strength. Therefore, the mass-to-flux ratios of the six clumps are approximately upper limits. All the six clumps are magnetically subcritical, indicating that the magnetic field is the dominant force, stronger than gravity. The enhanced magnetic field in the shell is strong enough to prevent these clumps from collapsing gravitationally, thereby reducing  the 
efficiency of star formation triggered by the expanding bubble. The case of the N4 bubble suggests that star formation triggered by expanding
bubbles might not easily occur and is likely to be highly dependent on the physical conditions of the local ISM. In this sense, 
it is expected that only very massive stars, which can sweep up sufficient molecular gas into shells via \ion{H}{2} region expansion and/or massive outflows, can overcome the resistance of magnetic fields to trigger star formation in shells. 

\acknowledgments
The authors acknowledge the anonymous referee's constructive comments. This work is supported by the Millimeter-Wave Radio Astronomy Database and the Key Laboratory for Radio Astronomy, CAS. Z.J. acknowledges support from NSFC 11233007. Z.C. acknowledges support from NSFC 11503087. A.R.L. acknowledges partial support from DIULS regular project PR15143. This research made use of NASA's Astrophysics Data System (Bibliographic Services).

\bibliographystyle{aasjournal}
\bibliography{myrefs}

\begin{thebibliography}{}
\expandafter\ifx\csname natexlab\endcsname\relax\def\natexlab#1{#1}\fi

\bibitem[{{Andersson} {et~al.}(2015){Andersson}, {Lazarian}, \&
  {Vaillancourt}}]{2015ARA&A..53..501A}
{Andersson}, B.-G., {Lazarian}, A., \& {Vaillancourt}, J.~E. 2015, \araa, 53,
  501

\bibitem[{{Beaumont} \& {Williams}(2010)}]{2010ApJ...709..791B}
{Beaumont}, C.~N., \& {Williams}, J.~P. 2010, \apj, 709, 791

\bibitem[{{Carpenter}(2001)}]{2001AJ....121.2851C}
{Carpenter}, J.~M. 2001, \aj, 121, 2851

\bibitem[{{Chandrasekhar} \& {Fermi}(1953)}]{1953ApJ...118..113C}
{Chandrasekhar}, S., \& {Fermi}, E. 1953, \apj, 118, 113

\bibitem[{Chen {et~al.}(2016)Chen, Zhang, Zhang, Jiang, Tamura, \&
  Kwon}]{2016ApJ...822..114C}
Chen, Z., Zhang, S., Zhang, M., {et~al.} 2016, The Astrophysical Journal, 822,
  114

\bibitem[{{Churchwell} {et~al.}(2006){Churchwell}, {Povich}, {Allen}, {Taylor},
  {Meade}, {Babler}, {Indebetouw}, {Watson}, {Whitney}, {Wolfire}, {Bania},
  {Benjamin}, {Clemens}, {Cohen}, {Cyganowski}, {Jackson}, {Kobulnicky},
  {Mathis}, {Mercer}, {Stolovy}, {Uzpen}, {Watson}, \&
  {Wolff}}]{2006ApJ...649..759C}
{Churchwell}, E., {Povich}, M.~S., {Allen}, D., {et~al.} 2006, \apj, 649, 759

\bibitem[{{Csengeri} {et~al.}(2014){Csengeri}, {Urquhart}, {Schuller}, {Motte},
  {Bontemps}, {Wyrowski}, {Menten}, {Bronfman}, {Beuther}, {Henning}, {Testi},
  {Zavagno}, \& {Walmsley}}]{2014A&A...565A..75C}
{Csengeri}, T., {Urquhart}, J.~S., {Schuller}, F., {et~al.} 2014, \aap, 565,
  A75

\bibitem[{{Deharveng} {et~al.}(2010){Deharveng}, {Schuller}, {Anderson},
  {Zavagno}, {Wyrowski}, {Menten}, {Bronfman}, {Testi}, {Walmsley}, \&
  {Wienen}}]{2010A&A...523A...6D}
{Deharveng}, L., {Schuller}, F., {Anderson}, L.~D., {et~al.} 2010, \aap, 523,
  A6

\bibitem[{{Elmegreen}(1998)}]{1998ASPC..148..150E}
{Elmegreen}, B.~G. 1998, in Astronomical Society of the Pacific Conference
  Series, Vol. 148, Origins, ed. C.~E. {Woodward}, J.~M. {Shull}, \& H.~A.
  {Thronson}, Jr., 150

\bibitem[{{G{\"u}ver} \& {{\"O}zel}(2009)}]{2009MNRAS.400.2050G}
{G{\"u}ver}, T., \& {{\"O}zel}, F. 2009, \mnras, 400, 2050

\bibitem[{{Krumholz} {et~al.}(2007){Krumholz}, {Stone}, \&
  {Gardiner}}]{2007ApJ...671..518K}
{Krumholz}, M.~R., {Stone}, J.~M., \& {Gardiner}, T.~A. 2007, \apj, 671, 518

\bibitem[{{Li} {et~al.}(2013){Li}, {Jiang}, {Liu}, \&
  {Wang}}]{2013RAA....13..921L}
{Li}, J.-Y., {Jiang}, Z.-B., {Liu}, Y., \& {Wang}, Y. 2013, Research in
  Astronomy and Astrophysics, 13, 921

\bibitem[{{Liu} {et~al.}(2016){Liu}, {Li}, {Wu}, {Yuan}, {Liu}, {Dubner},
  {Paron}, {Ortega}, {Molinari}, {Huang}, {Zavagno}, {Samal}, {Huang}, \&
  {Zhang}}]{2016ApJ...818...95L}
{Liu}, H.-L., {Li}, J.-Z., {Wu}, Y., {et~al.} 2016, \apj, 818, 95

\bibitem[{{Lockman}(1989)}]{1989ApJS...71..469L}
{Lockman}, F.~J. 1989, \apjs, 71, 469

\bibitem[{{Ostriker} {et~al.}(2001){Ostriker}, {Stone}, \&
  {Gammie}}]{2001ApJ...546..980O}
{Ostriker}, E.~C., {Stone}, J.~M., \& {Gammie}, C.~F. 2001, \apj, 546, 980

\bibitem[{{Pavel} \& {Clemens}(2012)}]{2012ApJ...760..150P}
{Pavel}, M.~D., \& {Clemens}, D.~P. 2012, \apj, 760, 150

\bibitem[{{Pecaut} \& {Mamajek}(2013)}]{2013ApJS..208....9P}
{Pecaut}, M.~J., \& {Mamajek}, E.~E. 2013, \apjs, 208, 9

\bibitem[{{Reid} {et~al.}(2016){Reid}, {Dame}, {Menten}, \&
  {Brunthaler}}]{2016ApJ...823...77R}
{Reid}, M.~J., {Dame}, T.~M., {Menten}, K.~M., \& {Brunthaler}, A. 2016, \apj,
  823, 77

\bibitem[{{Schuller} {et~al.}(2009){Schuller}, {Menten}, {Contreras},
  {Wyrowski}, {Schilke}, {Bronfman}, {Henning}, {Walmsley}, {Beuther},
  {Bontemps}, {Cesaroni}, {Deharveng}, {Garay}, {Herpin}, {Lefloch}, {Linz},
  {Mardones}, {Minier}, {Molinari}, {Motte}, {Nyman}, {Reveret}, {Risacher},
  {Russeil}, {Schneider}, {Testi}, {Troost}, {Vasyunina}, {Wienen}, {Zavagno},
  {Kovacs}, {Kreysa}, {Siringo}, \& {Wei{\ss}}}]{2009A&A...504..415S}
{Schuller}, F., {Menten}, K.~M., {Contreras}, Y., {et~al.} 2009, \aap, 504, 415

\bibitem[{{Shu} {et~al.}(1987){Shu}, {Adams}, \&
  {Lizano}}]{1987ARA&A..25...23S}
{Shu}, F.~H., {Adams}, F.~C., \& {Lizano}, S. 1987, \araa, 25, 23

\bibitem[{{van Marle} {et~al.}(2015){van Marle}, {Meliani}, \&
  {Marcowith}}]{2015A&A...584A..49V}
{van Marle}, A.~J., {Meliani}, Z., \& {Marcowith}, A. 2015, \aap, 584, A49

\bibitem[{{Wang} \& {Jiang}(2014)}]{2014ApJ...788L..12W}
{Wang}, S., \& {Jiang}, B.~W. 2014, \apjl, 788, L12

\bibitem[{{Watson} {et~al.}(2010){Watson}, {Hanspal}, \&
  {Mengistu}}]{2010ApJ...716.1478W}
{Watson}, C., {Hanspal}, U., \& {Mengistu}, A. 2010, \apj, 716, 1478

\bibitem[{{Wienen} {et~al.}(2012){Wienen}, {Wyrowski}, {Schuller}, {Menten},
  {Walmsley}, {Bronfman}, \& {Motte}}]{2012A&A...544A.146W}
{Wienen}, M., {Wyrowski}, F., {Schuller}, F., {et~al.} 2012, \aap, 544, A146

\bibitem[{{Yan} {et~al.}(2016){Yan}, {Xu}, {Zhang}, {Lu}, {Chen}, \&
  {Tang}}]{2016AJ....152..117Y}
{Yan}, Q.-Z., {Xu}, Y., {Zhang}, B., {et~al.} 2016, \aj, 152, 117

\end{thebibliography}

\end{document}